\documentclass[nofootinbib,superscriptaddress,twocolumn,showpacs,preprintnumbers,amsmath,amssymb,prl]{revtex4-1}

\usepackage{graphicx} 
\usepackage{grffile} 
\usepackage{slashed}
\usepackage{bbm} 
\usepackage{footmisc} 
\usepackage{multirow}

\usepackage{hyperref} 
\usepackage{amssymb} 
\usepackage{amsmath}
\usepackage{color}
\usepackage{xspace} 
\usepackage{caption}
\usepackage{subcaption}
\usepackage{natbib}

\usepackage{color,fancybox}
\usepackage{pifont}
\newcommand{\GeV}{\ensuremath{\,\mathrm{GeV}}\xspace}
\newcommand{\TeV}{\ensuremath{\,\mathrm{TeV}}\xspace}

\newcommand{\fb}{\ensuremath{\,\mathrm{fb}}\xspace}

\newcommand{\order}[1]{\mathcal{O}\!\left(#1\right)}

\newcommand{\bib}[1]{Ref.~\cite{#1}}
\newcommand{\fig}[1]{Fig.~\ref{#1}}

\begin{document} 

\title{$W\gamma$ production in vector boson fusion at NLO in QCD}

\preprint{FTUV-13-0930\;\; IFIC/13-67\;\; KA-TP-27-2013\;\;LPN13-063\;\;SFB/CPP-13-70}

\author{Francisco~Campanario} 
\email{francisco.campanario@ific.uv.es}
\affiliation{Theory Division, IFIC, University of Valencia-CSIC, E-46980 Paterna, Valencia, Spain.} 
\author{Nicolas Kaiser}
\email{nicolas.kaiser@student.kit.edu} 
\author{Dieter~Zeppenfeld}
\email{dieter.zeppenfeld@kit.edu} \affiliation{Institute for Theoretical Physics, KIT, 76128 Karlsruhe, Germany.}

\begin{abstract} 
The next-to-leading order QCD corrections to $W^\pm \gamma$ production
in association with two jets via vector boson fusion are calculated, including 
the leptonic decay of the W with full off-shell effects and spin
correlations. The process lends itself to a test of quartic gauge
couplings. The next-to-leading order corrections reduce
the scale uncertainty significantly and show a non-trivial phase space
dependence.
\end{abstract}

\pacs{12.38.Bx, 13.85.-t, 14.70.Fm, 14.70.Bh}

\maketitle


Di-boson production processes in association with two jets play an important
role at the LHC, not only as a background to searches for physics beyond the 
Standard Model, but also as a means to test the structure of the electroweak 
symmetry breaking sector. 
Within the Standard Model (SM), there are three distinct production modes at 
leading order~(LO). The QCD mechanism, i.e. the radiation of two partons 
in quark-antiquark annihilation to two vector bosons and crossing 
related processes, is of order $\order{\alpha_s^2\alpha^2}$ for 
on-shell production of both gauge bosons. 
For these processes, results at NLO QCD have been reported for 
$W^+W^-jj$~\cite{Melia:2011dw,Greiner:2012im}, 
$W^+W^+jj$~\cite{Melia:2010bm}, 
$W^{\pm}Zjj$~\cite{Campanario:2013qba},
including the leptonic decay of the vector bosons with all off-shell effects, and $\gamma\gamma jj$\cite{Gehrmann:2013bga} production.
In addition, there is the ``vector-boson-fusion''~(VBF) mechanism, which
is of order $\order{\alpha^4}$ at LO for on-shell production. The basic subprocess 
for the VBF channel is vector boson scattering, which means that the VBF 
processes are particularly interesting as a probe of electroweak symmetry breaking.
For weak boson scattering, the main focus will be on the scattering of longitudinal 
W's and Z's and the question, whether the recently discovered Higgs boson does indeed 
unitarize this process. However, electroweak boson scattering is also an 
excellent source of information on trilinear and quartic gauge couplings, and 
here, probing the scattering of transversely polarized gauge bosons is as 
important as the scattering of longitudinal modes. When considering 
transverse polarizations, final state photons are just as interesting as 
Z-bosons or W's. 
Finally, the production of three electroweak gauge bosons, with one off-shell 
gauge boson decaying into a quark-antiquark pair, is a third source of $VVjj$ events 
at order $\order{\alpha^4}$. NLO QCD corrections to $VVV$ production with 
leptonic decays are available via 
the {\texttt{VBFNLO}} program~\cite{Arnold:2008rz,*Arnold:2012xn}. 
Since the above three production modes peak in 
different regions of phase space, and because of their largely orthogonal 
color structures, interference between these modes is generally unimportant 
and can be neglected in most applications.

NLO QCD corrections for the VBF processes have been provided for all combinations 
of massive di-boson production in 
Refs.~\cite{Jager:2006zc,Jager:2006cp,Bozzi:2007ur,Jager:2009xx,Denner:2012dz}.
In this letter, we present the first theoretical prediction for the VBF production
of  $W^{\pm} \gamma jj$ final states at order $\order{\alpha_s\alpha^5}$. 
Compared to a massive gauge boson, which typically is observed in leptonic 
decays with a small branching fraction of order 
3 to 10~\%,
the production cross section  
of a final state on-shell photon is considerably enhanced.
%
In our calculation, the leptonic decay of the W boson is 
consistently included, with all
off-shell effects and spin correlations taken into account. This includes also
final state radiation, i.e. the radiative decay of the W.
Radiative W decays provide a sizable source of $\ell\nu\gamma jj$ events which 
diminish the sensitivity of $W\gamma jj$ production to anomalous couplings. 
In this letter, we also study how these contributions can be reduced safely.
In the following, we consider the specific leptonic final state 
$e^{\pm}\overset{\textbf{\fontsize{0.5pt}{0.5pt}\selectfont(---)}}{\nu_{e}}\gamma$. 
The final results can be multiplied by a factor two to take the  
$\mu^{\pm}\overset{\textbf{\fontsize{0.5pt}{0.5pt}\selectfont(---)}}{\nu_{\mu}}\gamma$ 
channel into account. 

This letter is organized as follows: After this introduction, we will
explain the major points of the implementation of $W^{\pm} \gamma jj$
production into the Monte-Carlo program 
{\texttt{VBFNLO}}~\cite{Arnold:2008rz,*Arnold:2012xn} and the checks
that we performed to assure its correctness. Then, the setup used for
the calculation and the numerical results for the cross
sections will be given. We will show that calculating the NLO QCD corrections reduces
the scale uncertainty and that these corrections have a non-trivial
phase space dependence. Finally, we will demonstrate how to suppress the
contributions from radiative W decay.
\begin{figure}[h] \centering
\includegraphics[width=0.65\columnwidth]{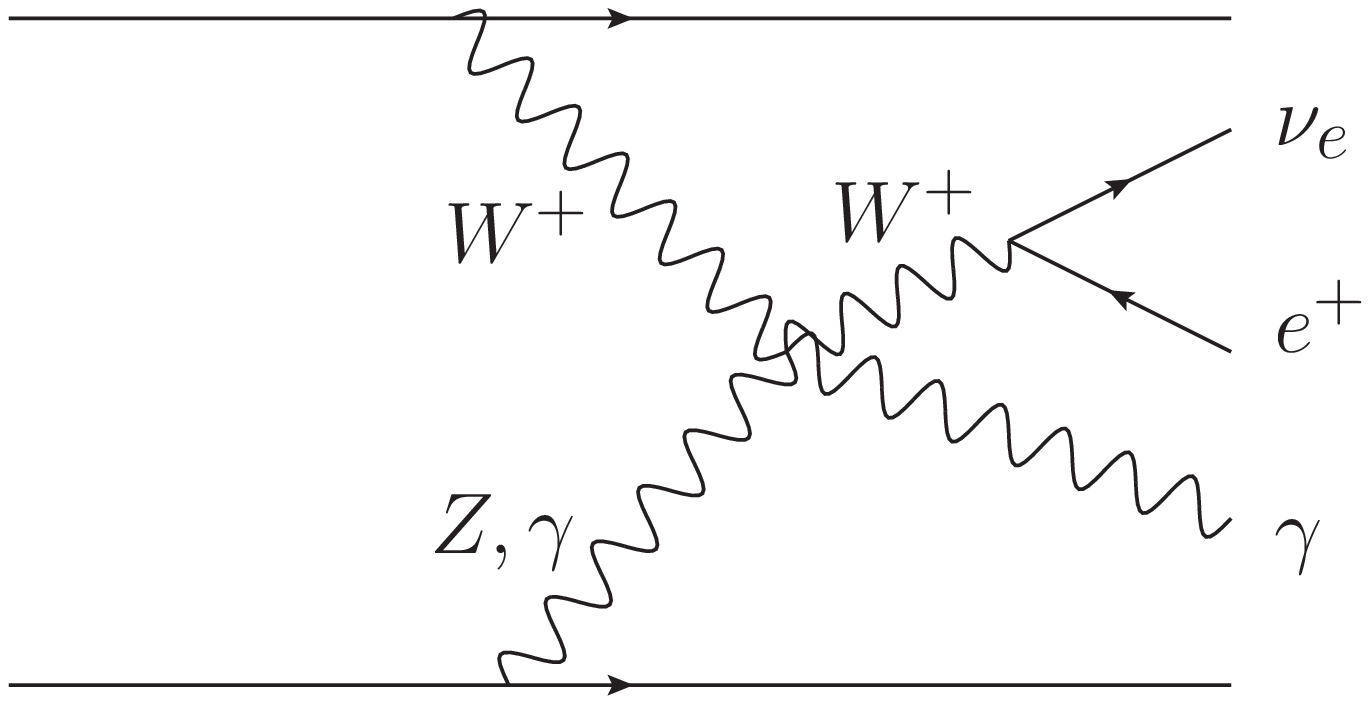}
\includegraphics[width=0.65\columnwidth]{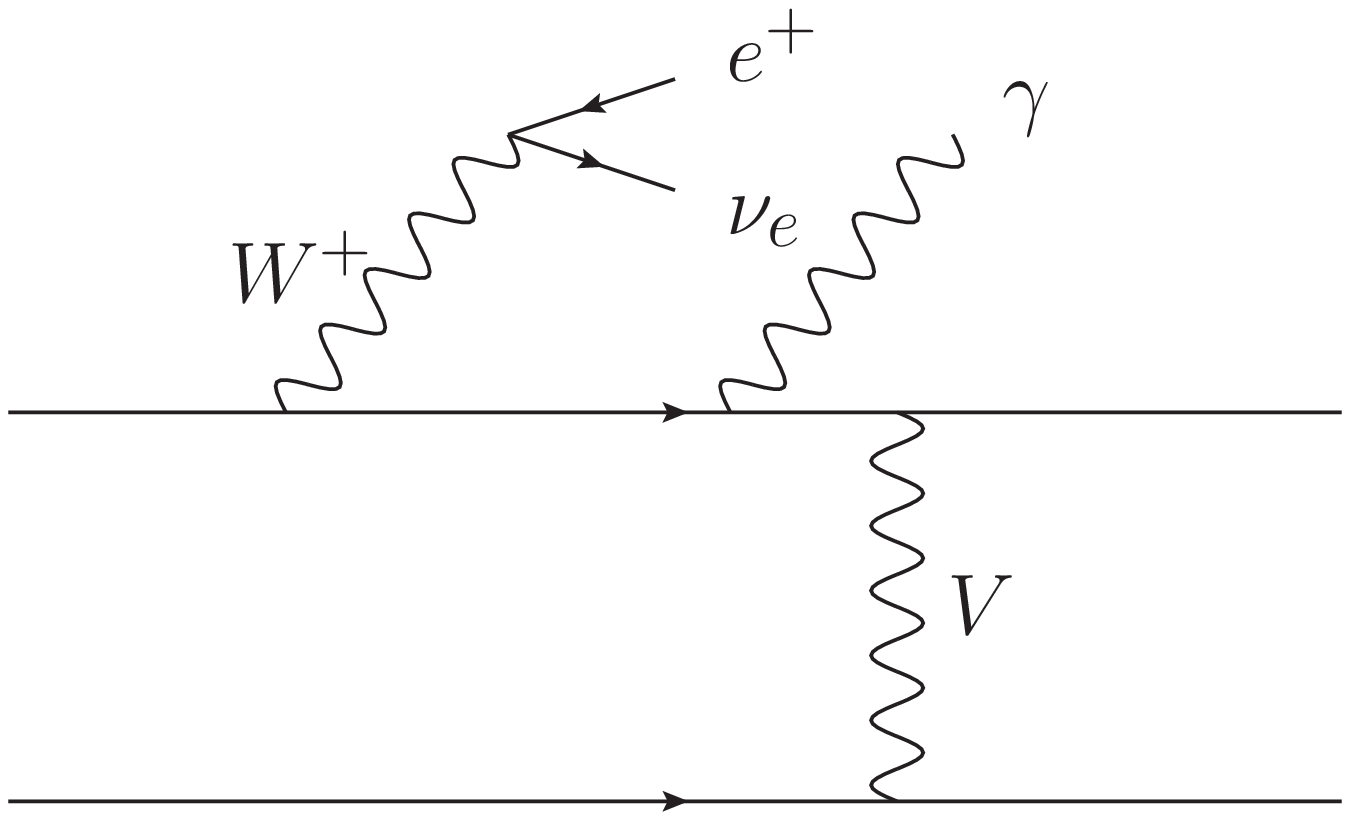}
\caption{Representative tree-level Feynman diagrams.}
\label{fig:feynTree}
\end{figure}
\section{Calculational Method}
Our calculation of the $W^\pm\gamma jj$ production cross section at NLO QCD closely
follows the one for $W^\pm Z jj$ production~~\cite{Bozzi:2007ur} 
and other VBF channels implemented in 
{\texttt{VBFNLO}}~\cite{Arnold:2008rz,*Arnold:2012xn}.
We consider only t-channel and u-channel Feynman diagrams but neglect interference 
between them. s-channel contributions, i.e. $VW\gamma$ production with subsequent 
hadronic decay of the (off-shell) $V$ are considered a separate process in 
{\texttt{VBFNLO}}. Contributions from s-channel diagrams as well as
interference terms between t-, u-, and s-channel contributions 
are strongly suppressed by the VBF cuts that we apply
in this letter.
For the calculation of the LO Feynman diagrams, e.g. the ones shown in
\fig{fig:feynTree}, the spinor-helicity formalism of
\bib{Hagiwara:1988pp} is used. 
The electroweak parts of the diagram are
combined to so-called leptonic tensors, which have to be calculated only
once per phase space point. For the calculation of the leptonic tensors,
we use the routines of the \textit{HELAS} package~\cite{Murayama:1992gi}.
The real emission~(RE) contributions comprise $q\to qg$
sub-diagrams, for example the ones that arise if one adds one gluon at 
every possible spot to the diagrams in~\fig{fig:feynTree}, as well as the
corresponding $g \rightarrow q\bar{q}$ diagrams. For the construction of
the RE diagrams, we use the same strategy as for the LO ones.
\begin{figure}[th]
  \centering
  \begin{subfigure}[b]{0.45\columnwidth} \centering
    \includegraphics[scale=0.32]{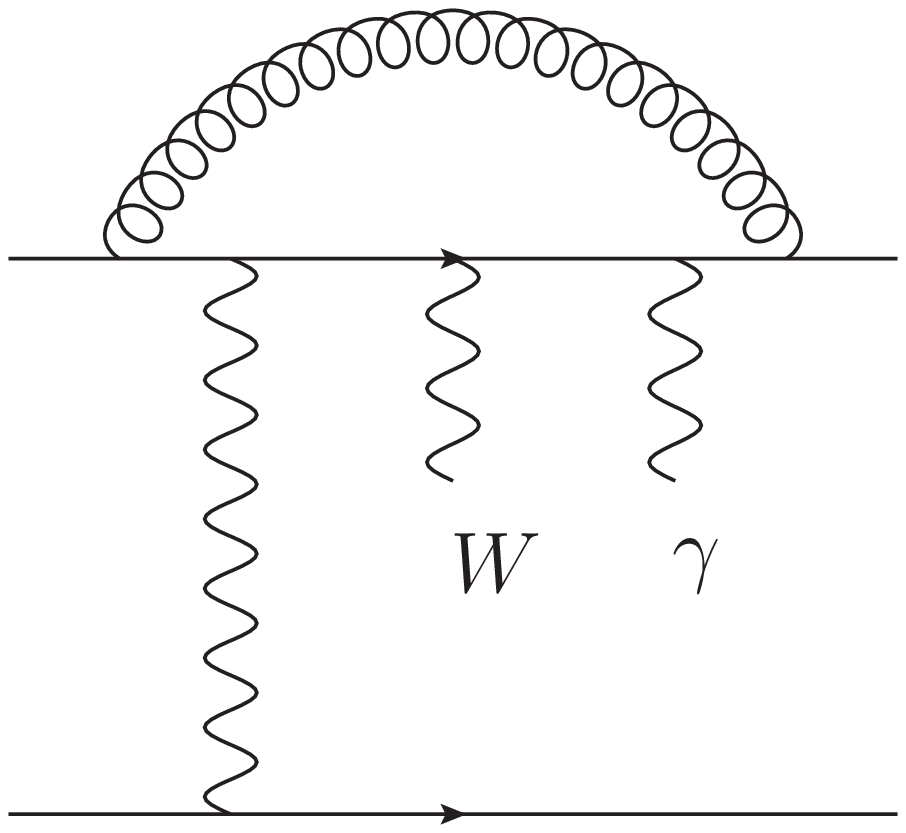}
    \caption{}
    \label{fig:Pentagon}
  \end{subfigure}
  \begin{subfigure}[b]{0.45\columnwidth} \centering
    \includegraphics[scale=0.32]{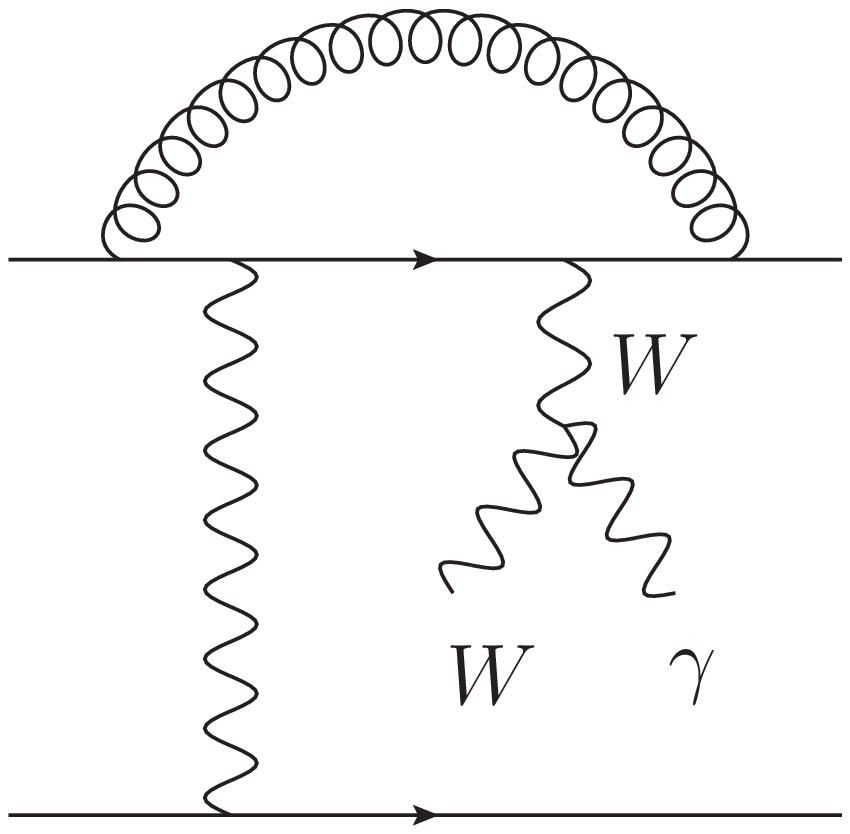}
    \caption{}
    \label{fig:WtoWABox}
  \end{subfigure}
  \begin{subfigure}[b]{0.45\columnwidth} \centering
    \includegraphics[scale=0.32]{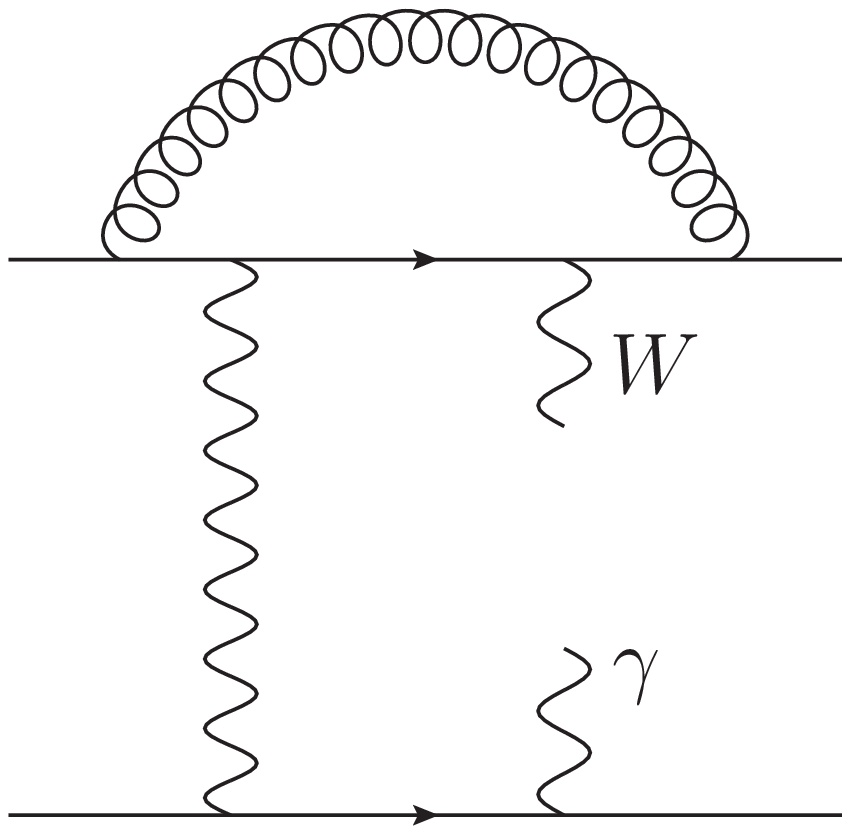}
    \caption{}
    \label{fig:Box}
  \end{subfigure}
  \begin{subfigure}[b]{0.45\columnwidth} \centering
    \includegraphics[scale=0.27]{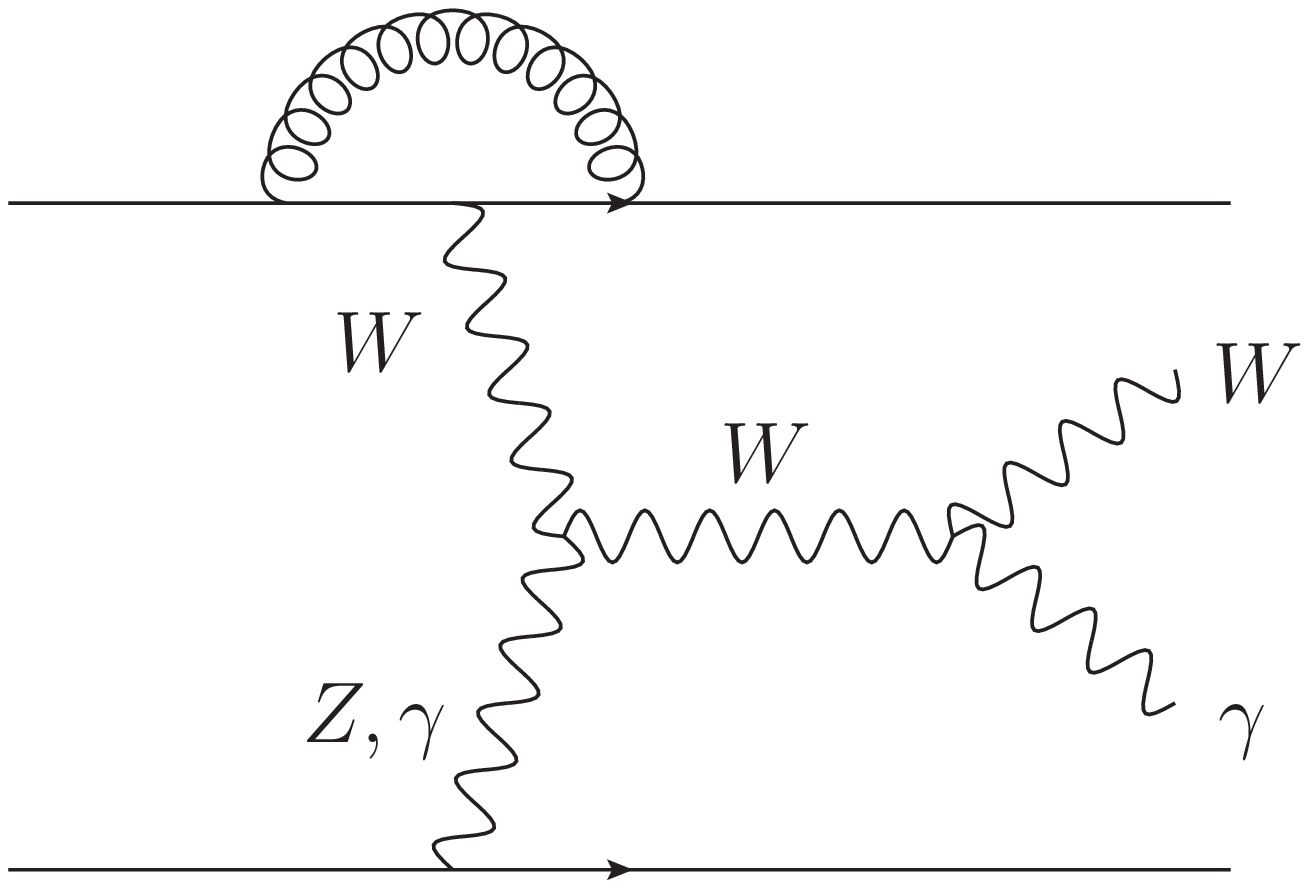}
    \caption{}
    \label{fig:Vertex}
  \end{subfigure}
  \caption{Selected Feynman diagrams contributing to the virtual amplitudes.}
  \label{fig:feynVirt}
\end{figure}
For the virtual amplitudes, we do not consider gluon exchange between the two 
quark lines. For the square of t-, u- or s-channel diagrams such contributions 
vanish due to color conservation. For interference terms, their 
contributions are also suppressed by the VBF cuts. 
Then, the virtual correction diagrams are obtained by adding a gluon loop
over each possible combination of the internal propagators and the
external legs of a single quark line of the LO diagrams.  
The lower diagram in~\fig{fig:feynTree} gets virtual corrections up to a pentagon on the
upper line as depicted in~\fig{fig:Pentagon}. Its lower line obtains
only vertex corrections similarly to the VBF Feynman diagram in
\fig{fig:Vertex}. In addition, there are several box diagrams, e.g. the
ones presented in~\fig{fig:WtoWABox} and~\fig{fig:Box}. The box
correction in the latter one can also be on the lower line.
All the virtual corrections to one quark line up to boxes or pentagons were calculated
using the \textit{Boxline} or \textit{Penline} routines of 
Ref.~\cite{Campanario:2011cs}, respectively.  For the regularization of the
infrared divergences, we use dimensional reduction 
and the Catani-Seymour dipole
subtraction method~\cite{Catani:1996jh} to make the virtual and the real
emission contributions numerically integrable in four dimensions. 
The subtraction procedure follows the one in \bib{Figy:2003nv} and,
hence, allows us to use an individual scale for each quark line.
We calculate the Born matrix elements, which are needed for the
calculation of the subtraction terms for $q\to qg$ and 
$\bar q\to \bar qg$ splitting, 
and cache them, so we can reuse them in the
$g \rightarrow q\bar{q}$ case.
Furthermore, we include anomalous couplings effects, which will
be studied in a forthcoming paper.
\\
For the potentially resonating $W^\pm$ and $Z$ propagators we use a variant 
of the complex mass scheme as implemented in 
\textsc{MadGraph}~\cite{Alwall:2007st}.
We checked our tree level matrix elements against
\textsc{MadGraph} and compared the cross sections to
\textsc{Sherpa}~\cite{Gleisberg:2008ta}. Applying VBF cuts, we found
complete agreement for the two jet cross sections but deviations of
1.5\% and 4\% for $W^{+}\gamma j j j$ and $W^{-}\gamma j j j$ production
respectively. We checked explicitly that these deviations originate from
the neglected s-channel contributions. They contribute at the per mille
level to the total NLO QCD corrections to $W^{\pm}\gamma j j $ production and,
therefore, can be safely neglected. 
The final state photon offers us the possibility to perform gauge tests
which are satisfied by all contributions.

Furthermore, the known~\cite{Campanario:2011cs} analytic expression of the poles 
for multiple vector boson emissions (on-shell or off-shell) along a quark line 
 has been compared numerically with the coefficients of the $1/\epsilon$ and
 $1/\epsilon^{2}$ poles 
computed in our process with the \textit{Boxline} and
\textit{Penline} routines. One typically finds agreement to
10 to 14 digits, 
thus, providing an additional strong check of the correctness of the implementation. %
%
In order to assure the numerical stability of the
calculation of the virtual contributions, we use Ward identity
checks. We set the amplitude to zero if the Ward
identities are satisfied with a precision worse than $\epsilon = 0.01$.
The share of phase space points, in which the Ward tests fail, is 0.36\%
for the \textit{Penline}s and $1.6\cdot10^{-6}\%$ for the
\textit{Boxline}s. Since the total NLO contributions are 
at the level of a few per cent, the error induced by setting the virtual 
amplitude to zero for those points is well below 
the statistical error, and, thus, negligible.

We checked the convergence of the dipoles of the Catani-Seymour subtraction. 
Moreover, we shifted parts of the terms proportional to 
$\left|{\cal M}_{\cal B}\right|^{2}$ from the five particle phase space
of the virtual corrections to the six particle phase space of the real
emission and found agreement within the numerical accuracy.

\section{Numerical Results}
\begin{figure}[t] \centering
  \scalebox{0.66}{\large \input{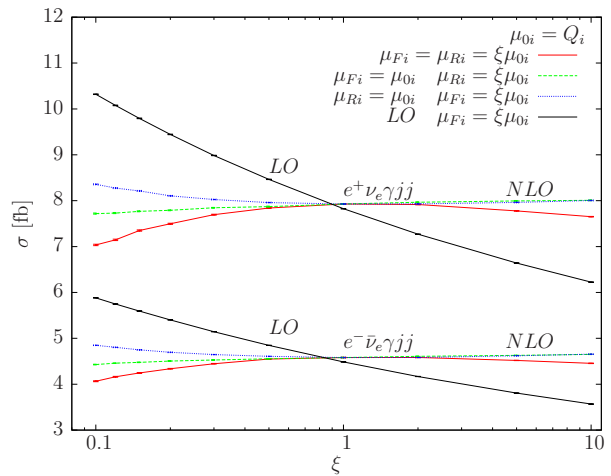}}
\caption{Scale dependence of the LO and NLO cross sections at the LHC. The
lower and upper curves stand for $pp \to e^- \bar{\nu}_e \gamma jj +X $ and 
$pp \to e^+ \nu_e \gamma jj +X $  production, respectively, 
at a c.m. energy of 14 TeV.}
\label{fig:scale}
\end{figure}

As EW input parameters, we use 
$M_Z=91.1876 \GeV$, $M_W=80.398 \GeV$ and $G_F=1.16637\times 10^{-5}\GeV^{-2}$ 
and derive the weak-mixing angle using the SM tree-level relations.  
All fermions aside from the top quark are considered to be massless.  
The width of the Z and the W bosons are calculated to be
$\Gamma_{Z}=2.508 \GeV$ and $\Gamma_{W}=2.098 \GeV$ respectively.
We use the CTEQ6L1 and CT10 parton distribution
functions (PDF)~\cite{Martin:2009iq} at LO and NLO respectively with 
\mbox{$\alpha_s^\text{LO}(M_Z)= 0.129808$} and
\mbox{$\alpha_s^\text{NLO}(M_Z)= 0.117982$}.  
The numerical results presented in this letter are calculated in 
the four-flavor scheme for the LHC at $14 \TeV$ center-of-mass energy. 
Effects from generation mixing are neglected~\cite{Bozzi:2007ur} 
since they almost completely cancel due to the unitarity of the CKM-matrix.
To reduce the contamination of s-channel contributions, we apply
typical VBF cuts. The charged lepton and photon are required to
be hard and central: $p_{T,\ell(\gamma)}\ge 20(30) \GeV$ and
$|y_{\ell(\gamma)}|\le 2.5$. 
%
Final state partons are clustered to jets using the anti-$k_t$
algorithm~\cite{Cacciari:2008gp} with the radius $R=0.4$. There must be
at least two hard jets with $p_{T,\text{jet}}\ge 30\GeV$ and $|y_{\text{jet}}|\le 4.5$.  
In addition, we impose a requirement on the lepton-jet and photon-jet
separation in the azimuthal angle-rapidity plane 
$\Delta R_{\ell,j} \ge 0.4$ and $\Delta R_{\gamma,j} \ge 0.7$, where
only jets passing the above cuts are considered. 
Furthermore, we use the photon isolation criterion \`{a} la Frixione~\cite{Frixione:1998jh} with a cone radius of $\delta_0=0.7$,
efficiency $\epsilon=1$ and exponent $n=1$.
The VBF cuts, which we apply, are 
\mbox{$\left|y_{j_{1}}-y_{j_{2}} \right| > 4$}, \, 
\mbox{$y_{j_{1}} \times y_{j_{2}} < 0$} and 
\mbox{$m_{j_{1} \, j_{2}} > 600  \GeV$}. 
%
As the central value for the factorization and renormalization scales, we
choose $\mu_{Fi}=\mu_{Ri}=\mu_{0}= Q_i $, where $Q_i$
is the absolute value of the momentum transferred from quark line $i$ to
the EW process.
With this setup, we obtain  
$\sigma_{LO} = 7.828 \, \pm \, 0.005 \fb \,
( 4.486 \, \pm \,0.003 \fb )$
and 
$\sigma_{NLO} = 7.910  \, \pm \, 0.007 \fb \,
(4.588 \, \pm \, 0.005 \fb )$ for 
$W^{+}\gamma jj \, (W^{-}\gamma jj)$ production, with the $W$ decaying into 
the first generation of leptons.
The K-factor, defined as $  K\equiv \sigma_{NLO}/\sigma_{LO}$, is
$1.013 \, (1.021)$. \\
Since we calculate only a fixed order in perturbative QCD, our results
depend on two unphysical scales, the factorization scale $\mu_{Fi}$ and the
renormalization scale $\mu_{Ri}$.
The scale variation plot in Fig.~\ref{fig:scale} shows that the scale
dependence of both processes can be significantly reduced by calculating
the NLO QCD corrections.

\begin{figure}[t] \centering
  \scalebox{0.65}{\large \input{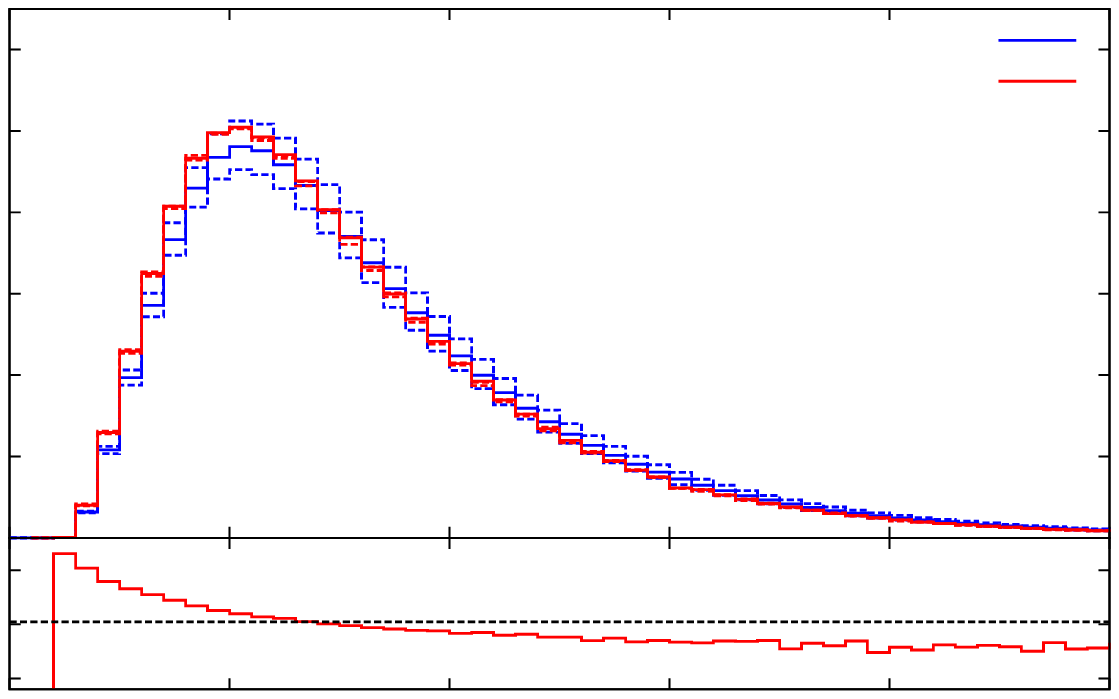}}
  \scalebox{0.65}{\large \input{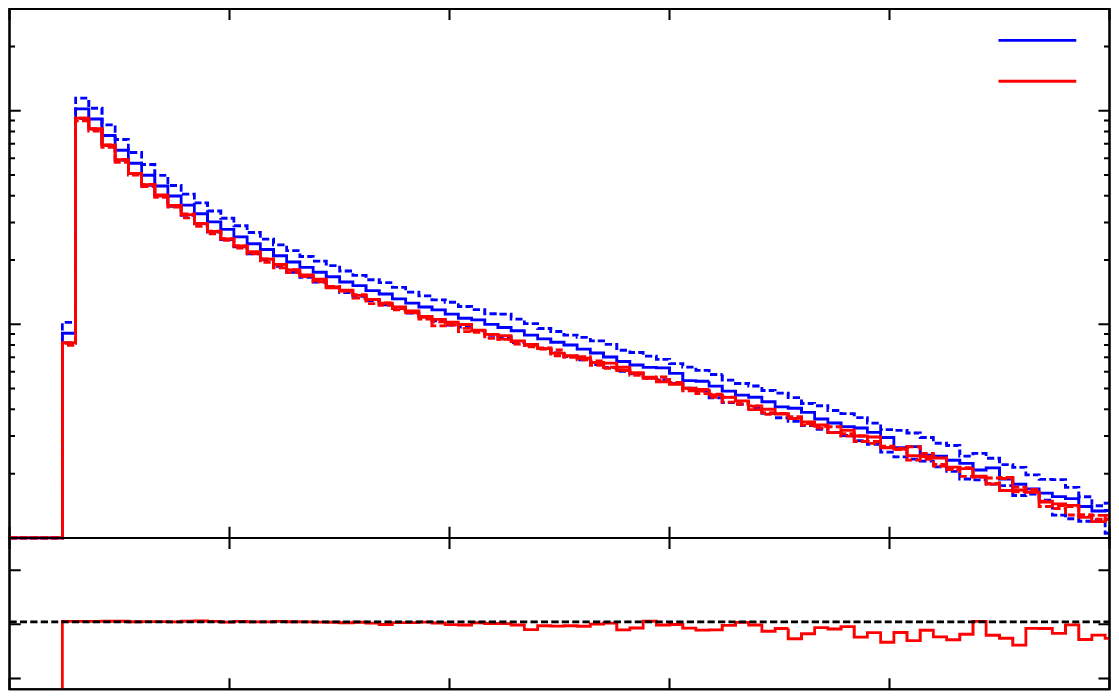}}
  \caption{Differential cross sections and K-factors for the transverse
momentum of the hardest jet (top) and the photon (bottom). The dashed
lines indicate the results for $\mu_{Fi}=\mu_{Ri}=\mu_{0i}= 2^{\pm 1}\, Q_i  $.
}
\label{fig:dist_pt}
\end{figure}

\begin{figure}[h] 
  \centering
  \scalebox{0.78}{\large \input{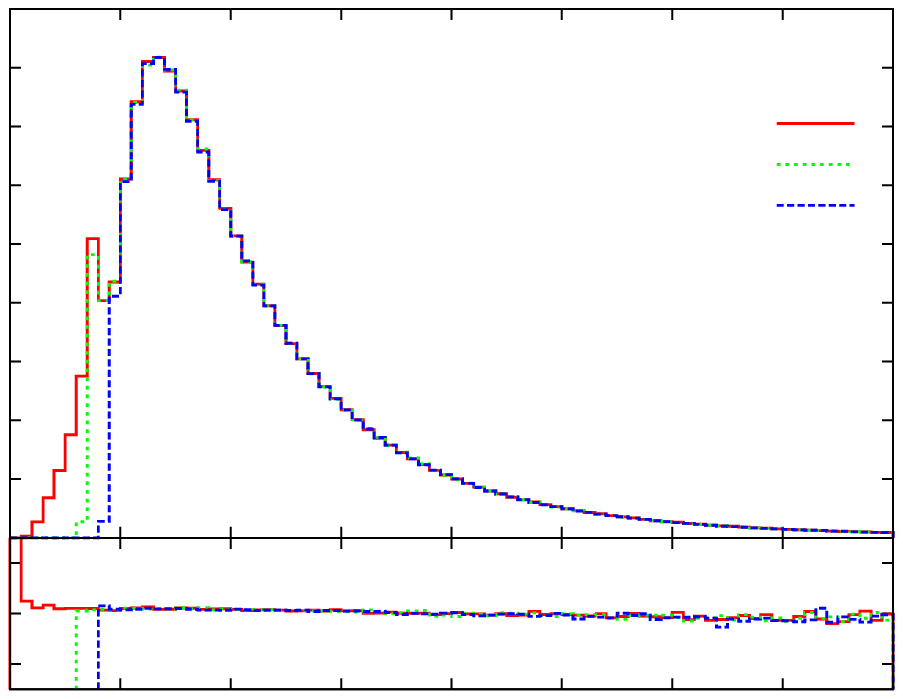}}
  \scalebox{0.78}{\large \input{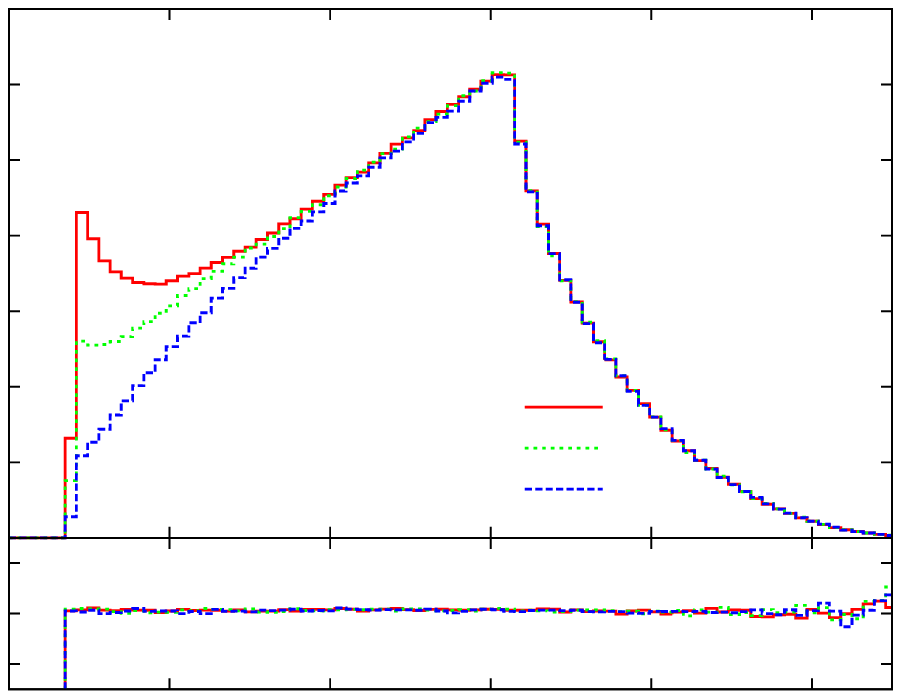}}
  \caption{Differential distribution of the transverse mass of the
    two gauge bosons $m_{T,W\gamma}$~(upper plot) and differential lepton-photon
    R separation distribution~(lower plot) for different values of the
    $m_{T,W\gamma}$ cut. Lower panels show the corresponding
    differential K-factors.}
\label{fig:finstatrad}
\end{figure}

In the following, distributions for the 
$W^+ \gamma j j$ production channel will be presented. 
Fig.~\ref{fig:dist_pt} shows the differential LO and NLO cross sections
over the transverse momentum of the hardest jet (upper plot) and the
photon (lower plot) as well as the differential and the total K
factors. To give a measure for the scale uncertainty, we also plot the
results for $\mu_{Fi}=\mu_{Ri}=\mu_{0i}= 2^{\pm 1} Q_i$
(dashed lines). In the $p_{T,j_1}$ and the $p_{T,\gamma}$ distributions,
the relative scale uncertainty is approximately constant over the whole
range examined. In both cases, it can be significantly reduced by
calculating the NLO QCD results. While the differential K-factor in the
$p_{T,\gamma}$ distribution is stable over the whole range examined, in
the $p_{T,j_1}$ distribution it decreases continuously over the whole
range.\\ 
In processes with at least one lepton and one photon in the final state,
the photon can be radiated off the lepton. 
This radiative W decay represents a simple QED process,
which diminishes the sensitivity to anomalous couplings in our case. 
In order to suppress radiative W decay, we follow \bib{Baur:1993ir} and 
define the transverse cluster mass of the $W\gamma$
system as 
$
 m_{T,W\gamma} = \left(\left[(m_{l\gamma}^2 +
      p_{T_{l\gamma}}^{2})^{\frac{1}{2}} + \displaystyle{\not}p_{T} \right]^2 -
    (  {\boldsymbol p_{\boldsymbol T} }_{l\gamma} + { \displaystyle{\not} \boldsymbol p_{\boldsymbol T} } )^{2} \right)^{\frac{1}{2}}.
$
The upper panel in \fig{fig:finstatrad} shows the corresponding distribution.
By imposing the cut $m_{T,W\gamma}>90 \GeV$,
the radiative decay peak at $m_{T,W\gamma}=m_{W}$ can be removed. This
cut reduces the final state radiation contributions significantly and affects
mainly the region of small $R_{l\gamma}$~(lower panel). Additionally, the NLO
cross section is only reduced by approximately 10\%, which shows the efficiency of the cut.

In this letter, we have presented the first calculation 
of $W^{\pm} \gamma jj + X$ production in VBF at order 
$\order{\alpha_s \alpha^5}$. The factorization and scale
uncertainties at NLO are significantly reduced and one finds K-factors close
to one. 
We plan to make the code publicly available as part of 
the {\texttt{VBFNLO}} program~\cite{Arnold:2008rz,*Arnold:2012xn}.

We acknowledge the support from the Deutsche Forschungsgemeinschaft via the
Sonderforschungsbereich/Transregio SFB/TR-9 Computational Particle Physics.
FC is funded by a Marie Curie fellowship (PIEF-GA-2011-298960) and partially
by MINECO (FPA2011-23596) and by LHCPhenonet (PITN-GA-2010-264564).  

\bibliographystyle{h-physrev} \bibliography{EWWAjj_short}

\end{document}